\documentclass[english,american]{article}
\usepackage[latin9]{inputenc}
\usepackage{geometry}
\usepackage{color}
\usepackage{verbatim}
\usepackage{amsthm}
\usepackage{amsmath}
\usepackage{amssymb}
\usepackage{graphicx}
\usepackage{esint}

\makeatletter
\theoremstyle{plain}
\newtheorem{thm}{\protect\theoremname}
\theoremstyle{remark}
\newtheorem{rem}[thm]{\protect\remarkname}

\usepackage{amsfonts}
\usepackage{epsfig}\usepackage[american]{babel}
\usepackage{graphicx}

\providecommand{\remarkname}{Remark}
\providecommand{\theoremname}{Theorem}

\providecommand{\remarkname}{Remark}
\providecommand{\theoremname}{Theorem}

\makeatother

\usepackage{babel}
\addto\captionsamerican{\renewcommand{\remarkname}{Remark}}
\addto\captionsamerican{\renewcommand{\theoremname}{Theorem}}
\addto\captionsenglish{\renewcommand{\remarkname}{Remark}}
\addto\captionsenglish{\renewcommand{\theoremname}{Theorem}}
\providecommand{\remarkname}{Remark}
\providecommand{\theoremname}{Theorem}

\begin{document}

\title{Cyber-Physical Modeling and Control of Crowd of Pedestrians: A Review
and New Framework}

\author{Ke-cai~Cao%
\thanks{caokc@njupt.edu.cn%
},\\
College of Automation, Nanjing University of Posts and Telecommunications,
\\
Nanjing, P.R. China, 210023.\\
YangQuan Chen %
\thanks{ychen53@ucmerced.edu%
},\\
Mechatronics Embedded Systems and Automation Lab, School of Engineering,
\\
University of California, Merced, CA, USA 95343.\\
Dan Stuart %
\thanks{idahoeinstein@gmail.com%
},\\
Department of Electrical and Computer Engineering, \\
Utah State University, Logan, UT, USA 84322., \\
Dong Yue,\\
College of Automation, Nanjing University of Posts and Telecommunications,\\
 Nanjing, P.R. China, 210023.}
\maketitle
\begin{abstract}
Recent advances in modeling and control of crowds of pedestrians are
briefl\textcolor{black}{y surveyed in this paper. Possibilities of
applying fractional calculus in the modeling of crowd of pedestrians
have been shortly reviewed and discussed from different aspects such
as descriptions of motion, interactions of long range and effects
of memory. Control of the crowd of pedestrians have also been formulated
using the framework of Cyber-Physical Systems and been realized using
networked Segways with onboard emergency response} personnels to regulate
the velocity and flux of the crowd. Platform for verification of the
theoretical results are also provided in this paper.

\end{abstract}

\section{Introduction}

As the most socially complex animals on the planet, research related
to crowd of pedestrians has received a lot of attention in recent
years. A lot of work has been conducted from perspectives of behavior,
psychology, cognition and network to analyze problems related to particles,
vehicles, robots, animal and even human beings (See \cite{Moussaid2011},
\cite{Czirok1999}, \cite{Spieser2009a}, \cite{Couzin2008}, \cite{Couzin2005}).
Some manuscripts have been published in recent years concerning the
modeling and control problem of crowd of pedestrians such as \cite{AlfioQuarteroni2003},
\cite{Haken2006}, \cite{Thalmann2007}, \cite{WolframW.F2008}, \cite{Bellomo2008a},
\cite{Kachroo2008b}, \cite{Pelechano2008}, \cite{Timmermans2009},
\cite{Kachroo2009}, \cite{Barcelo2010}.

On the other side, catastrophic events occurring around the world
have demonstrated the need to re-analyze and re-examine current evacuation
policies and procedures for crowd of pedestrians. The dynamic and
uncertain nature of disasters has required the need of changing backup
contingency plans according to evacuation needs. The first problem
that confronted in the research of crowd of pedestrians is how to
obtain a satisfactory model to characterize the complex nature of
this kind of dynamic systems. \textcolor{black}{There are a lot of
characteristics which should be considered in the modeling and control
problems such as self-organization, following leaders, common motives
for action, psychological unity, emotional intensity, level of violence
as shown in \cite{Berlonghi1995}.} Due the features mentioned above,
modeling and control of crowd of pedestrians are challenging tasks
as shown in recent work of \cite{DirkHelbing2005}, \cite{Bellomo2011a},
\cite{Cristiani2011}, \cite{ChristianDogbe2012}, \cite{Bogdan2011a}
and \cite{DanielStuart2013}.

Recent studies on modeling and control of crowd of pedestrians have
been firstly reviewed in this paper in three different scales. Then
necessities of introducing fractional calculus in the modeling and
control of crowd of pedestrians have been discussed in detail and
networked Segways with on-board emergency response personnels have
been employed in the evacuation control of crowd of pedestrians. Framework
based on Cyber-Physical systems have been formulated in the end for
studying the modeling and control problem of crowd of pedestrians.
Main contributions of this paper lie in that the tool of fractional
order calculus have been introduced in the modeling and control of
crowd of pedestrians where some phenomena observed from realities
are impossible or not easy to be considered just in the framework
of integer order calculus. The aim of this paper is to provide an
alternative way for studying the complex crowd of pedestrians which
is much more close to reality.

The rest of the paper is organized as follows. Model for crowd of
pedestrians are firstly reviewed in three different scales in Section
\ref{sec:Model} respectively; Then recent advances in control of
crowd of pedestrians are discussed in micro-scale and macro-scale
in Section \ref{sec:Control}; Our framework for studying the evacuation
problem of crowd of pedestrians are shown in Section \ref{sec:Our-Framework}
based upon calculus of fractional order. Validation of obtained theoretical
results are also included in this Section; Conclusions of this paper
are given in Section \ref{sec:Conclusion}.

\section{Modeling of Crowd of Pedestrians\label{sec:Model}}

A lot of modeling methods have been \textcolor{black}{proposed} recently.
Most of these modeling methods are based on inspirations that the
crowd of pedestrians can be treated as moving particles, smoothing
fluids, or granular flows in different scales. Modeling methodologies
are firstly reviewed in micro-scale, macro-scale and mesco-scale,
respectively. Then necessity of introducing fractional order calculus
in the modeling of crowd of pedestrians are discussed and some recent
advances of modeling using fractional order calculus are also included
in this section.

\subsection{Micro-scale---Ordinary Differential Equation (ODE)}

When the density of pedestrians is low, each pedestrian can move freely
and interactions among pedestrians can be modeled using the framework
of social forces. A lot of research has also been conducted using
the Newton's laws as shown in the following microscopic model
\begin{equation}
m_{i}\frac{dv_{i}}{dt}=f_{i}^{S}+\sum_{j=1}^{n}f_{ij}^{N}+\sum f_{k}^{W},\label{eq:micro model}
\end{equation}
where $f_{i}^{S}$ is the self-driven force towards some desired velocity,
$f_{ij}^{N}$ is the interaction between agent $i$ and its neighbor
$j$ and $f_{k}^{W}$ represents the interactions with environment
such as walls or corridors.

Different $f_{ij}^{N}$ and $f_{k}^{W}$ have been constructed in
the work of \cite{Helbing2000} for panic scenarios and it has been
proved that neither internal forces nor external forces alone can
perform well in the evacuation of crowd of pedestrians and the practical
way is to consider them simultaneously in the micro-scale model (\ref{eq:micro model});
Similarly, it has been shown in \cite{Lohner2010} that not only external
forces such as interactions with neighbours and environments but also
some internal forces such as will force and personal information should
be considered in the modeling process. But the fact is that a lot
of work has been done from the view of multi-agent system where only
external forces are considered. Interactions with external environment
such as the spatial-temporal markings have been described as potential
function and added into the right-side of (\ref{eq:micro model})
as done in \cite{Helbing1998}; Relationship between formation patterns,
stability and types of interactions have also been analyzed in \cite{Orsogna2006}
under external potential function to find whether a swarm will collapse
or not.\textcolor{black}{{} Some recent work such as \cite{YiquanSong2013},
\cite{Rodriguez2012} from the aspect of modeling and \cite{Bing-ChangWang2012},
\cite{TaoLi2008} from the aspect of control of crowd of pedestrians
are also only focused on external interactions with neighbouring agents. }

Most of the simulation results have been conducted in this scale as
shown in \cite{Pelechano2008} because the methods are simple and
animations are realistic. In order to satisfy the requirements of
high realism and real-time animation in the modeling and control in
this scale, the idea of mapping desired behaviors to the stable solutions
of classical nonlinear dynamic system such as Van der Pol oscillator
or fix-point attractor has been introduced in \cite{MartinA.Giese2009}
and \cite{Mukovskiy2013} where\textcolor{black}{{} one nonlinear transformation
connecting pedestrian's periodic or non-periodic motion with structurally
stable system has been firstly constructed }and then real-time animation
problem can be solved in low-dimensional space without losing details;
Methods from cognitive science such as heuristic behavior have been
employed to adjust the walking direction and speed in micro-scale
in \cite{Moussaid2011}; Global roadmap-based navigation method has
been proposed in the homing or exploring problem as shown in \cite{O.BurchanBayazit2004}.
Although these methods are simple and efficient, the prescription
of all behaviors is not very easy for crowd of pedestrians of large
numbers.
\begin{rem}
Comments on Microscopic Model:
\begin{itemize}
\item \textcolor{black}{Main advantages of the microscopic model lie in
that heterogeneities of each pedestrian can be considered explicitly
and simulation results obtained are of highly realistic. But the microscopic
model is not a good choice if the number of pedestrians are very high
as some unnecessary interactions or effects have been included; }
\item \textcolor{black}{Most of the previous work has just treated each
pedestrian as one physical particle and few works have been done under
the consideration of pedestrian's memory or some other internal effects;
Local interacting rules have received a lot of attentions in previous
research while long range interacting rules are not so popular. }
\end{itemize}
\end{rem}

\subsection{Macro-scale---Partial Differential Equation (PDE)}

The motion of crowd of pedestrians show some striking analogies with
the motion of fluids when the density goes high. Thus research of
crowd of pedestrians in macro-scale has benefited a lot from the macro-scale
research of traffic system where the well known LWR (Lighthill\textendash Whitham
and Richards) model (\ref{eq:LWR}) and PW (Payne-Whitham) model (\ref{eq:PW})
have been proposed. It is well known that the traffic dynamic system
in macro-scale as shown in \cite{Kachroo2009} can be described by
\begin{equation}
\frac{\partial}{\partial t}\rho(t,x)+\frac{\partial}{\partial x}f(t,x)=0\label{eq:LWR}
\end{equation}
and
\begin{align}
\frac{\partial}{\partial t}\rho(t,x)\!\!+\!\!\frac{\partial}{\partial x}\rho(t,x)v(t,x)\!\! & \!=0\label{eq:PW}\\
\frac{\partial}{\partial t}v(t,x)\!\!+\!\! v(t,x)\frac{\partial}{\partial x}v(t,x)\!\! & \!=\frac{V(\rho)\!-\! v}{\tau}\!\!-\!\!\frac{A(\rho)_{x}}{\rho}\!\!+\!\!\mu\frac{v_{xx}}{\rho}\nonumber
\end{align}
where $\rho$ denotes the density of crowd and $f(t,x)$ is the flux
of crowd, $V(\rho)$ is equilibrium Speed, $\frac{V(\rho)-v}{\tau}$
is the relaxation term, $\frac{A(\rho)_{x}}{\rho}$ is the anticipation
term and $\mu\frac{v_{xx}}{\rho}$ is the viscosity term. Relationship
between flux and density has attracted a lot of interests and a lot
of work has been done to generalize the LWR model and PW model.

Two versions of LWR model such as high-order LWR model and low-order
LWR model have been shown in \cite{Zhang2001}. Compared with classical
LWR model that only concerns transitions among equilibrium states
and homogeneity, both non-equilibrium transitions and inhomogeneities
have been accommodated in the work of \cite{Zhang2001} through constructing
much more complex relationship between flux and density; Similarly,
new relationship between speed and density has been used to derive
Lighthill-Whitham model in panic scenarios in \cite{Colombo2010}.

Also based on the fundamental conservation law of mass and momentum,
macroscopic model for crowd pedestrians in two dimensional space,
as below
\begin{equation}
\frac{\partial}{\partial t}\rho+\frac{\partial}{\partial x}(\rho u)+\frac{\partial}{\partial y}(\rho y)=0\label{eq:Roger model}
\end{equation}
has been constructed in \cite{RogerL.Hughes2002} for different types
of pedestrians with different walking habits. Macroscopic model in
two dimensional space has been generalized in \cite{Jiang2010a} with
terms for anticipation and relaxation.

As interactions among pedestrians are assumed to be the same in macro-scale
therefore it is not easy to incorporate the heterogeneity of each
pedestrian. In order to characterize the crowd of pedestrians more
precisely, models that can benefit from both the micro-scale and macro-scale
are much preferred. A modeling procedure based on the time-varying
measures
\begin{equation}
\mu_{t}=\theta m_{t}+(1-\theta)M_{t},\label{eq:measure}
\end{equation}
has been prescribed for the multi-scale characterization of crowd-pedestrian
system where $m_{t}=\sum_{j=1}^{N}\delta P_{j}(t)$ and $dM_{t}(x)=\rho(t,x)dx$
are the microscopic and macroscopic mass respectively. Both topological
interactions such as the Braess's paradox (obstacles may contribute
to improve the flow of people in some situations) characterized in
macro scale and anisotropic interactions such as the granular role
of some pedestrians described in micro-scale can also be considered
in this framework through the time-varying measure (\ref{eq:measure})
; Similar results that are composed of agent-based microscopic model
and flow-based macroscopic model have been shown in \cite{Xiong2009}
and \cite{Xiong2010} where the accuracy of microscopic model and
the efficiency of macroscopic model have been combined together through
initializing each other by techniques of aggregation and disaggregation.
\begin{rem}
Comments on Macroscopic Model:
\begin{itemize}
\item \textcolor{black}{Computation time have been greatly decreased as
each pedestrian has been treated as physical particle and have same
characteristics in macro-scale; }
\item \textcolor{black}{Main disadvantages of macroscopic models lie in
that heterogeneities of pedestrians such as interactions or mobilities
cannot be characterized or considered in detail in this scale. Obtained
results using the macroscopic model just provide some references for
the control of crowd of pedestrians. }
\end{itemize}
\end{rem}

\subsection{Mesco-scale---Integral Differential Equation (IDE)}

After Boltzmann formalized the concepts of kinetic equations in the
nineteenth century, it has been widely used in astrophysics, engineering,
social science and even in biology. Although the following basic kinetic-transport
equation

\begin{equation}
\frac{\partial}{\partial t}f(t,x,\xi)+\xi\cdot\nabla_{x}f(t,x,\xi)=0\label{eq:mesco equation}
\end{equation}
that describes the evolution of density $f(t,x,\xi)$ of particles
passing through $(t,x)$ with velocity $\xi$ seems very simple, great
success has been achieved in the applications of Boltzmann's kinetic
equation and many generalizations of this equation have been made.

A lot of modeling results for\textcolor{black}{{} crowd of pedestrians}
have been derived from the equation (\ref{eq:mesco equation}) since
macro-scale variables and micro-scale variables have combined with
each other in this equation. Nonlinear integral-differential equations
have also been used in \cite{Arlotti2000} to describe the \textcolor{black}{crowd
of pedestrians} with competitions and kinetic interactions as

\[
\frac{\partial f_{i}}{\partial t}(t,u)=J_{i}[f](t,u)+\gamma_{i}(t,u)
\]
where the $J_{i}[f](t,u)$ describes the evolution of density due
to gain and loss of distribution function and $\gamma_{i}(t,u)$ describes
the production and migration of the group; Mescoscopic models with
binary interactions, averaged binary interactions and mean field interactions
have been given in \cite{Bellomo2002} as

\begin{equation}
\frac{\partial f}{\partial t}+\xi\frac{\partial f}{\partial x}+\frac{\partial(fF[f])}{\partial\xi}=Q(f,u)\label{eq:Bellomo}
\end{equation}
where different $F[f]$ and $Q(f,u)$ are employed to describe these
interactions. Conservation law of mass and momentum in macro-scale
has been generalized in \cite{Helbing1998c} where internal intentions
and external interactions have been simultaneously considered in the
following mescoscopic model
\begin{align}
\frac{\partial}{\partial t}\rho_{\mu}+\frac{\partial}{\partial x}(\rho_{\mu}v_{\mu}) & =\!\int m_{\mu}q_{\mu}\! dv_{\mu}\nonumber \\
 & +\!\sum_{\mu}[\frac{m_{\mu}}{m_{v}}\rho_{v}\chi_{\mu}^{\mu v}(1)\!-\!\rho_{\mu}\chi_{\mu}^{\mu v}(1)]\label{eq:fluid helbing}
\end{align}
where $\rho_{\mu}$ is the density, the first term on the right is
caused by pedestrians entering or leaving some interested areas and
the second term describes the effects caused by internal intentions
and external interactions. Equation (\ref{eq:fluid helbing}) has
been generalized to the following two dimensional mescoscopic equation

\begin{align*}
\partial_{t}\rho+[\partial_{x_{1}}(\rho v_{1})+\partial_{x_{2}}(\rho v_{2})] & +[\partial_{v_{1}}(\rho A_{1})+\partial_{v_{2}}(\rho A_{2})]\\
 & =(\partial_{t}\rho)_{event}^{+}-(\partial_{t}\rho)_{event}^{-}
\end{align*}
in \cite{SergeP.Hoogendoorn2007} where $\partial_{x_{1}}(\rho v_{1})+\partial_{x_{2}}(\rho v_{2})$
means the changes of density due to convection, $\partial_{v_{1}}(\rho A_{1})+\partial_{v_{2}}(\rho A_{2})$
are the terms of acceleration and deceleration, $(\partial_{t}\rho)_{event}^{+}-(\partial_{t}\rho)_{event}^{-}$
means the interaction of events.

Some other research have also been conducted from the aspect of micro-scale
using the framework of mesco kinetic theory. With the help of activity
variables, heterogeneities of each pedestrian have been characterized
in micro-scale in\textbf{ }\cite{Lillo2007}\textbf{ }and\textbf{
}\cite{Bellomo2011} using the statistical distribution of position
and velocity; Enskog-like interactions and stochastic interactions
have been considered in \cite{Delital2003}; Nonlinear interactions
instead of linear interactions have been applied in \cite{Arlotti2012};
Both short and long range interactions have been reported in \cite{Lillo2007}
contrary to previous modeling methods where only local interactions
are assumed.
\begin{rem}
Comments on mescoscopic model:
\begin{itemize}
\item As both information from micro-scale and information from macro-scale
have been included in this framework, heterogeneities of pedestrians
and different kinds of interactions can be included easily in the
right hand side of (\ref{eq:Bellomo}); Another advantage is that
efficiency of the mescoscopic model has been greatly improved compared
with that of microscopic model;
\item \textcolor{black}{One of the main disadvantages of mescoscopic model
is that existence of analytic expressions for equilibrium is not guaranteed
for obtained IDEs now and another disadvantages of this framework
is the technical difficulty when there are more than one microscopic
variables that needed to be included in the above IDEs; }
\end{itemize}
\end{rem}

\subsection{Fractional Model}

Fractional Calculus has shown great potential in different applications
such as particles in fluid, plasma physics, quantum optics and many
others. Some phenomena such as self-similarity, non-stationary and
spiky phenomena, non-Gaussian relaxation, short or long memory and
long range interactions are all closely related to fractional calculus.
The authors have noticed that previous research has mentioned or implied
potential applications of fractional calculus in modeling, control
and optimization of \textcolor{black}{crowd of pedestrians} in different
aspects such as description of motion, description of interactions
and hysteresis phenomenona. Recent advances in these aspects are shortly
reviewed in the followings.
\begin{enumerate}
\item Fractional Calculus in Description of Motion

Some applications of fractional calculus has been done for physical
particles in macro scale. Based on the fractional derivative of order
$\alpha$ in \cite{Hilfer2000}

\begin{equation}
\frac{\partial^{\alpha}f(x,t)}{\partial t^{\alpha}}=\underset{\triangle t\rightarrow0}{lim}\frac{f(x,t+\triangle t)-f(x,t)}{\triangle t^{\alpha}},\label{eq:definition fractal order}
\end{equation}
some different fractional kinetic equations (FKE) and their properties
have been explained in \cite{G.M.2002} such as the FKE with desired
direction, anisotropic FKE and nonlinear FKE; Then fractional kinetics
of distributed-order has been proposed in \cite{Sokolov2004} and
it has been shown that the fractional model with distributed order
can be effectively used to describe the accelerating and decelerating
of sub-diffusion and super-diffusion.

Based on these preliminary works, fractional model for crowd of pedestrians
in the macro-scale has been proposed in \cite{Ke-CaiCao2012} using
the fractal derivatives of time and space where fractional order dynamics
for one dimensional crowd can be described as

\begin{equation}
\frac{\partial}{\partial t^{\alpha}}\rho(t,x)+\frac{\partial}{\partial x^{\beta}}[\rho(t,x)v(t,x)]=0\label{eq:my frac model}
\end{equation}

where $\rho\text{(\ensuremath{t,x})}$ and $v(t,x)$ are the density
and velocity of crowd at position $x$ and time $t$, $\alpha$ and
$\beta$ is the fractal order of time and space respectively. Similar
results can also be found in \cite{Long-FeiWang2014} where $\alpha=\beta$
has been assumed. Considering that each pedestrian's decision making
process plays an important role in his (her) next movement, the dynamic
process of decision making has been described in \cite{Bogdan2011a}
where coupling dynamic models composed of the Fractional Fokker-Planck
equation

\begin{equation}
\frac{\partial^{\alpha}P(u,t)}{\partial t^{\alpha}}=-\frac{\partial[f(u,d,t)P(u,t)]}{\partial u}+\Xi(P,u,t)\label{eq:forward}
\end{equation}
and the Fractional Master equation

\begin{equation}
\frac{\partial^{\alpha}u(x,y,t)}{\partial t^{\alpha}}=f(u,d,t)+g(u,d,t)\xi(t)+\eta(t)\label{eq:backward}
\end{equation}
have been constructed and $P(u,t)$ is the distribution function,
$\Xi(P,u,t)$ is the additional terms due to variations in infinitesimal
time increment, $u(x,y,t)$ defines the cost that each pedestrian
should pay for desired destinations, $f(u,d,t)$ is the nonlinear
dependence on previous costs $u(x,y,t)$ and previous decisions $d(t)$.
The forward dynamics (\ref{eq:forward}) describes the evolution of
distribution function with respect to time while the backward equation
(\ref{eq:backward}) describes the evolution of pedestrian's decision
with respect to the inverse of time. In order to control and predict
the evolution of crowd of pedestrians, optimal control of this complex
system has been firstly reformulated in \cite{Bogdan2011} based on
the fractional master equations.

Concerning the fractional model in micro-scale, fractional Langevin
equations have been given in \cite{M.Romanovas2009} for human motion
tracking and fractional models with fixed order and variable order
have been proposed in \cite{Romanovas2012} and \cite{Romanovas2013}
to describe motion of pedestrians.

\item Long Range Interactions of Power Law

Different formation patterns for swarming of fish can be generated
through tuning the area of attraction and repulsion as shown in \cite{Couzin2002};
Quantitatively characterizing of relationship between the area of
repulsion and attraction have been analyzed in \cite{Mogilner2003}
for dynamic model of integer order. Thus even for dynamic models of
integer order, phenomena of transiting from ordered state to disordered
state have been observed in \cite{Toner1995} under long range interactions.
Transition from chaos to turbulence has been reported in \cite{Zaslavsky2007}
where fractional long range interactions of $\frac{1}{l^{1+\alpha}}$
($l$ is the distance between oscillators, $\alpha$ is the tuning
parameter) has been imposed for nonlinear oscillators. Recently, transitions
have also been declared for Boltzmann equations using local, non-local
or configuration-dependent interactions in \cite{Max-OlivierHongler2014}
where transition from purely diffusive regime to flocking patterns
can be realized by modulating the range and strength of interactions.
We can say that range of interaction and power of interaction play
an important role in study of system of large numbers.

Actually, the long range interactions are connected with fractional
dynamics. Relationships between long range interactions of power law
in micro-scale and fractional Euler-Lagrange equations in macro-scale
have been discussed in \cite{Ishiwata2012} and it has been proved
that there exists one fractional Euler-Lagrange equation in macro-scale
if there are long range interactions of power law in micro-scale;
From which we can say that the gap between micro-scale and macro-scale
can be bridged together using long range interactions of power laws.
As the long range interactions are so common in crowd of pedestrians,
it is reasonable to describe the dynamics of crowd of pedestrians
using fractional calculus.

\item Hysteresis phenomenon and Collective Memory

It has been shown in \cite{Couzin2002} that minor changes in individual's
responses can lead to different collective behaviors such as alignment,
swarm and torus where the collective memory has played an important
role; Hysteresis phenomenon has been observed in \cite{Couzin2008}
where nonlinear relationship between collective behaviors and range
of interactions has been explicitly shown in Figure 1 of \cite{Couzin2008}.

According to the following Definition of Hysteresis from Wikipedia:\textbf{\textit{\textcolor{black}{{}
Hysteresis}}}\textit{\textcolor{black}{{} is the dependence of the output
of a system not only on its }}\textbf{\textit{\textcolor{black}{current
input}}}\textit{\textcolor{black}{, but also on }}\textbf{\textit{\textcolor{black}{its
history of past inputs}}}\textit{\textcolor{black}{. }}In other words,
the change of group's behavior not only depends on the current control
input but also depends on the history of individual behavior and the
shape of group; The hysteresis phenomenons or effects of short memory
are so common that they can be found in the acceleration, deceleration
and equilibrium of traffic flows in \cite{H.M.1999} and individual's
actions such as buying and selling in the stock market. For particles
with long time memory, fractional Fokker-Planck equations have been
derived in \cite{Tarasov2008} based on the correlating functions
of probability densities that follow power law; A non-negative memory
function $\eta(\cdot)$ has been introduced in \cite{Colombo2012}
to generalize the LWR model using the form of convolution
\[
\rho\star\eta=\int_{R^{2}}\eta(x-\xi)\rho(t,\xi)d\xi
\]
where preferred path can be found and regions of high density can
be avoided.

\end{enumerate}

\textcolor{black}{As calculus of integer order can be considered as
special case of fractional order calculus, each pedestrian can be
characterized much closer to reality using the framework of fractional
order calculus. Many effects such as individual's memory, long-range
interactions which are difficult to characterize using calculus of
integer order can be compensated using calculus of fractional order; }

\begin{rem}
Compared with dynamic models of integer order in previous research,
some advantages and disadvantages of dynamic model of fractional order
are listed in the followings:
\begin{itemize}
\item \textbf{\textcolor{black}{In time domain:}}\textcolor{black}{{} Only
normal diffusive process has been considered in previous study in
macro scale due to the limitation of calculus of integer order. Besides
the normal diffusive process, sub-diffusive process and super-diffusive
process have been added to describe the crowd in macro scale using
fractional calculus in time domain; }
\item \textbf{\textcolor{black}{In spatial domain:}}\textcolor{black}{{} Dimension
of space are only limited to 1, 2 or 3 in previous research while
the fractional dimension can be included in fractional calculus to
consider effects of environments on crowd pedestrian system; }
\item \textbf{\textcolor{black}{Some other elements:}}\textcolor{black}{{}
As we have shown that some phenomenons such as self-similarity, non-stationary
and spiky phenomenons, non-Gaussian relaxation, effects of memory,
long-range interactions which are difficult to characterize or explain
using calculus of integer order can be compensated using calculus
of fractional order; }
\end{itemize}

The authors do not want to prove that previous models of integer order
are not effective anymore in reality and we just want show that Fractional
Calculus as generalization of Calculus of Integer order has provided
us much more freedom in characterizing and understanding of the complexities
of crowd of pedestrians. The authors admit that there are a lot of
challenging work left to do for the obtained model of fractional order
such as controller design, stability analysis and performance evaluation.
Most of them are not so easy to solve now.

\end{rem}

\section{Control of Crowd of Pedestrians\label{sec:Control}}

\subsection{Control of Macroscopic Model }

In the evacuation of crowd-pedestrian system, moving direction and
moving speed are of great importance to guarantee the smooth movement
of crowds. Some previous research has chosen $v(t,x)$ in (\ref{eq:PW})
as the control input and distributed state feedback controllers have
been constructed in \cite{Wadoo2006} and \cite{Wadoo2006a} for the
macro-scale model in one and two dimensional space. Main idea of \cite{Wadoo2006}
and \cite{Wadoo2006a} is using finite dimensional system to approximate
infinite dimensional system; Then theory of nonlinear control can
be used for design of feedback controllers for this approximating
system. Main problems of this framework is that the original system
may be still unstable even if the obtained controllers work very well
on the approximating system. Different from the framework of approximation,
the control and stability problem of the macroscopic model (\ref{eq:PW})
has been formulated directly in the distributed control of partial
differential equations in \cite{Wadoo2008} and \cite{Wadoo2010}.
Previous controllers have been generalized to diffusion, advective
and advective-diffusion controllers. Comparisons of these controllers
have been done in \cite{Dong2013} and it has been shown that much
more serious problems may arise if only diffusion feedback controller
is adopted as there is no preferred direction for the pedestrians
to follow in evacuation process; But faster evacuation can be realized
for LWR model under diffusion-advection-state feedback controllers
since the direction of evacuation is provided.

In order to avoid undesirable congestion and blockages in the evacuation
of pedestrians, optimal feedback controllers have been proposed in
\cite{Shende2011} and \cite{Shende2013} by instructing pedestrians
to adjust their velocities. Although optimal results have been obtained
in \cite{Shende2011} and \cite{Shende2013}, applications of these
results are only limited to the deterministic case. Robust control
of crowd-pedestrian system has interested a lot of people in recent
years. For example, Lyapunov techniques have been firstly utilized
in \cite{Alvarez1999} to construct velocity controllers for automated
highway systems where not only position and time but also effects
of lanes, drivers and destinations have been included in the macroscopic
PDE model; Previous diffusion-state feedback controllers have been
generalized using methods of Lyapunov redesign in \cite{Yang2013}
to deal with disturbances in control input; Robust controllers based
on sliding model control have been constructed in \cite{Wadoo2013}
for the control of crowd of pedestrians with matched and unmatched
uncertainties that are caused by external disturbance and parametric
variations.
\begin{rem}
Similar to the modeling methods for macroscopic model, heterogeneities
of controllers have also been neglected in this scale.
\end{rem}

\subsection{Control of Microscopic Model}

\subsubsection{Control using Leaders with Naive followers}

Based on inspirations from swarming of ants, schooling of fish and
flocking of birds, how to formulate or control the collective behaviors
in micro-scale has received a lot of attentions from the community
of control, computation and computer. One of the fascinating phenomenon
is that this kind of collective behaviors can be generated from simple
or basic interacting rules and control of this kind of collective
behaviors can be realized through controlling just a small part of
the group.

Determining of moving direction based on neighbor\textquoteright s
direction have been firstly proposed in \cite{AliJadbabaie2003} for
average consensus problem of each agent. Work of \cite{Couzin2005}
and \cite{Couzin2008} has explicitly shown the quantitative relationship
between individual's local interacting rules and group's collective
behaviors; Three experiments have been done in \cite{JohnR.GDyer2009}
to show the effects of number and topology of informed agents on the
consensus problem and it has been proved using experiments that only
a small number of informed individuals are enough for driving a large
number of uninformed individuals to reach consensus without explicit
communications; Further results concerning consensus under uncertain
or conflicting information have been considered in \cite{JohnR.G.Dyer2008}.
For evacuation problem of crowd pedestrian system, result of \cite{Francois2004}
has shown that the evacuation rate can be greatly improved by adding
some leaders with global knowledge and it is always good for the evacuation
process if we place some leaders in the immediate proximate of crowds
and some leaders scattered around the environment.
\begin{rem}
Similar to the pinning control of complex network with large number
of nodes, how many nodes are needed to control the network and which
one should be chosen as the pinning node are interesting problems.
For crowds, how many leaders and what kind of leaders are needed in
control of crowd of pedestrians are worthy of further considerations.

\textcolor{black}{Phenomena such as self-similar structure, heavy-tailed
workloads and long-range dependence effects that have been proposed
in \cite{KihongPark2000}, \cite{PaulBogdan2011a} for study of network
are related to calculus of fractional order. The authors think that
not only topology and connectivity should be considered but also the
dynamic evolutions of the network itself should be considered especially
when confronted with stochastic network of large numbers Thus control
in the framework of fractional order calculus is very interesting
although there are many challenges. }
\end{rem}

\subsubsection{Control without leaders}
\begin{itemize}
\item \textbf{Decentralized Framework: }

Using the methodology of decomposition, control problems for complex
system can be transformed into control problems for simpler subsystems.
A lot of decentralized controllers have been obtained under the framework
of decomposition. Decentralized controllers for large number of stochastic
agents have been given in \cite{Li2008a} where not only the evolution
in time scale but also the evolvement in ``space'' scale ( where
$N\rightarrow\infty$, $N$ is the number of agents) are considered;
Decentralized controllers have also been constructed using the framework
of decomposition in \cite{Huang2010b} where complex LQG games problem
has been reduced to two-player games problem.

Advantages of the decentralized framework are that the burden of computation
has been greatly reduced and \textcolor{black}{obtained results can
be easily extended to systems with large numbers. However, neighbor\textquoteright s
information is not used in the decentralized framework, where it is
much preferred to have distributed controllers.}

\item \textbf{Distributed Framework: }

Some other distributed feedback controllers for macroscopic model
have been reviewed in previous sections or can be found in \cite{Wadoo2006a},
\cite{Wadoo2008} and \cite{Wadoo2010}. Some consensus protocols
of multi-agent system also belong to this framework which we will
not state one by one and interested readers can find some recent advances
in the review papers of \cite{YongcanCao2013},\cite{Hai-BoMin2012}
or references in them.

One big challenge for distributed control lies in the burden of computation
if the number of pedestrians goes to infinity. Some previous research
has adopted the mean field methods to estimate the influence of neighboring
agents to relieve the burden of computation. Mean field methods have
been used in \cite{Nourian2011} for controller design for leader-follower
agents with linear stochastic dynamics; Distributed controllers for
agents with one major/leader agent have been given in \cite{Wang2012a}
with Markov jump parameters in controlled system and random parameters
in objective functions where mean field method has been utilized to
estimate effects from neighboring pedestrians and the leader; Mean
field LQG controllers have been proposed in \cite{Huang2012a} for
socially optimal control problem whose cost functions are coupled
with each other; Mean field control strategies have been given in
\cite{Nourian2013} for the consensus problem of multi-stochastic
agents based on the assumption that information of initial state distribution
is available to each agent; Similar results can also be found in \cite{Nourian2012a}
in the control of stochastic multi-agent systems where statistical
information of neighboring agents are obtained using mean field methods.

In order to study the true interacting rules among agents, game theory
has also been combined with mean field method in control of agents
of large numbers which we will call it mean field games in the followings.
Based on mean field games introduced in \cite{LASRY2007}, dynamics
of human's decision making process has been considered in \cite{Christian2010}
where mean field game theory has been employed in the backward Hamilton-Jacobi-Bellman
(HJB) equation and forward Fokker-Plank (FP) equation; The Forward-backward
equations have also been adopted in \cite{Lachapelle2011} to describe
the aversion and congestion phenomena in macroscopic scale where each
pedestrian's ability of anticipation has been characterized using
the backward equation;

\end{itemize}
\textcolor{black}{Advantages of the mean field method is that only
information of initial distributions is needed in the design of controllers
and the popular topology condition such as connected graph or jointly-connected
graph is no longer needed; Additional benefit is that communications
among neighboring individuals are no longer needed and obtained results
can also be easily extended to systems with large numbers. }
\begin{rem}
Research of crowd of pedestrians has benefit a lot from the research
of traffic control system where limiting speed has served as an effective
way to guarantee the normal flow of vehicles. Information of density
has been used in specifying limitation of speeds in different area
in \cite{Kachroo2009}. Roadmap-based planning has been used in\cite{Rodriguez2012}
for evacuation of large number of agents where interactions and coordination
among agents are of great importance in realization of the evacuation.

Speed, density and even interactions are the main information that
has received a lot of attention from the view of multi agent system.
Besides the above information, it is the authors' belief that effects
of memory, environment or even building structure should be considered
in modeling and control of crowd of pedestrians where Fractional Calculus
will play an important rule.
\end{rem}

\section{Our Framework\label{sec:Our-Framework}}

A fractional framework for modeling and controlling of crowd of pedestrians
has been proposed in this paper. The proposed research represents
the coupling of the `PHYSICAL PART' (Mass Pedestrian Evacuation management
of crowds) with the `CYBER PART' (modeling and prediction of crowds).
And transferring of information between these two parts has been implemented
through networked Segways, with on-board emergency response personnels,
and facility sensing and actuation.
\begin{figure*}[tbh]
\begin{centering}
\includegraphics[scale=0.6]{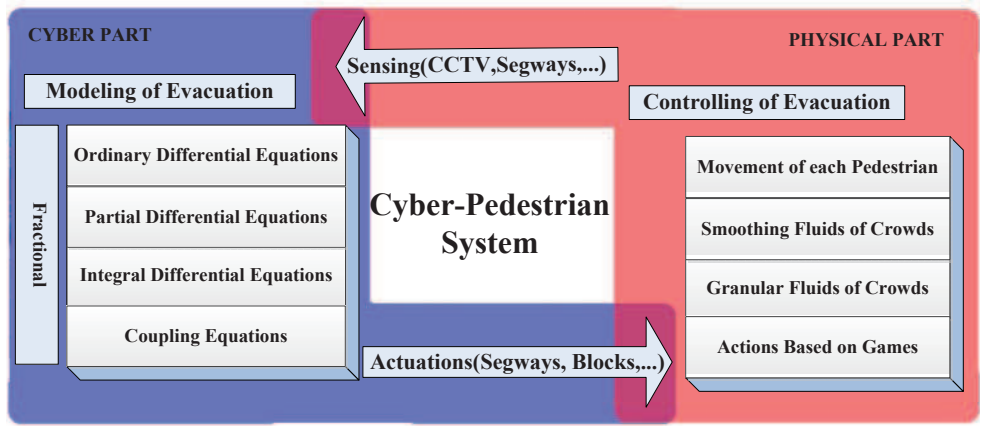}
\par\end{centering}

\noindent \centering\protect\protect\caption{\selectlanguage{english}%
\textit{\label{fig:CPS}}\foreignlanguage{american}{\textit{A Cyber
Physical Framework for describing the coupling of model, control,
and management}}\selectlanguage{american}%
}
\end{figure*}

\textcolor{black}{In Cyber Part, ordinary differential equations,
partial differential equations and integral differential equations
have been employed to describe the crowd of pedestrians using calculus
of fractional order in micro-scale, macro-scale and meso-scale, respectively.
Interesting information such as speed, density, flux and even formation
patterns that obtained through CCTV, Segways, Cellphone and some other
sensors can be used in calculating the obtained models, controlling
the crowds and even predicting the stampede that is going to occur.
Closed loop system that are composed by the cyber part and physical
part in Figure \ref{fig:CPS} has shown the flow of information and
major points of each part.}

\subsection{Fractional Modeling of Crowd of Pedestrians }
\begin{itemize}
\item In Micro-scale level with low density of crowds, behavior of each
pedestrian can be described by the ordinary differential equations
based on some widely used methods such as social force model (\ref{eq:micro model})
or agent based model where parameters of systems can be further calibrated
using empirical or observed data. Main reason of using these classical
models is due to the convenience of introducing heterogeneities in
the description of pedestrians in micro-scale;
\item In Macro-scale level, the density of crowds is so high that the motion
of all pedestrians can be modeled as continuum fluids where partial
differential equations can be derived from the conservation law of
mass or momentum on interested area. Main differences between crowd
of pedestrians and smoothing fluids are that different motion patterns
such as crossing or intersecting is allowed in crowd-pedestrian system
due to the freedom of choosing different routes. The authors believe
that both the time scale and the spatial scale should be considered
in the modeling of crowd of pedestrians and preliminary work of fractional
macroscopic model (\ref{eq:my frac model}) has been shown in \cite{Ke-CaiCao2012}
where the fractional order associated with time and the fractional
order associated with the fractal structure have been included;
\item In Mesco-scale level with medium density, the dynamics of evacuation
or egress process is similar to the diffusion process of active particles
in many aspects such as the porous or granular pattern in the smoothing
fluids. Fractional convection-diffusion equations are useful tools
to model the crowd pedestrians in this case due to the phenomenon
of porosity observed in this scale; Heterogeneous pedestrians can
be modeled using mobile potential fields indexed by \textcolor{black}{activity
variables} as shown in Figure \ref{fig:potential field} which can
be included in the convection-diffusion equations to guarantee the
existence of heterogeneities on this level; Interactions between microscopic
model and macroscopic model can also be realized through mean field
games to increase the validation of obtained models and relieve the
burden of computation.

\begin{figure}[tbh]
\centering\includegraphics[scale=0.12]{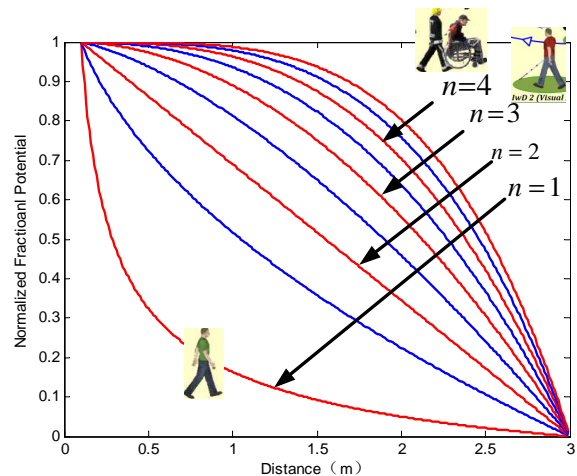}\protect\protect\caption{\selectlanguage{english}%
\label{fig:potential field}\foreignlanguage{american}{Normalized
effect of activity variables on potential fields (PFs) }\selectlanguage{american}%
}
\end{figure}

\end{itemize}
\textbf{Reasons of using the above modeling framework: }
\begin{itemize}
\item As there is no general method prescribed for modeling crowd pedestrians
for all kinds of scenarios, it is much reasonable to choose the most
appropriate model for different problems. Since the microscopic model
is powerful to describe the heterogeneities of pedestrians, we choose
to use the microscopic model when the density is low; With the increasing
of density, granular flows with porosity phenomenon can be observed
in large crowds. This kind of heterogeneities are modeled using different
mobile potential fields which can be easily added into the right hand
side of (\ref{eq:Bellomo}) to describe the effects from micro-scale;
As density grows, the porosity or granular phenomenon of smoothing
fluids disappears and fractional dynamic model can also be obtained
by using generalized conservation law of mass or momentum as deduced
in previous research.
\item The macroscopic model is responsible for generating a homogenizing
effect with desirable smoothness in macro-scale, and the microscopic
model is responsible for characterizing heterogeneities and interactions
from the macro-scale. For mescoscopic model, not only heterogeneities
and porous patterns of crowds can be explicitly characterized but
also interactions between micro-scale and macro-scale can be included
in the obtained integral differential equations.
\item Although we describe the modeling of crowds in different scales according
to their densities, these obtained models are not independent of each
other as shown in Figure \ref{fig:Model-fractional}. The macro-scale
variables such as density or flow comes from aggregation of micro-scale
data of each pedestrian and the motion of each pedestrian is also
affected or constrained by the people around him.

\begin{center}
\begin{figure*}[tbh]
\begin{centering}
\includegraphics[scale=0.4]{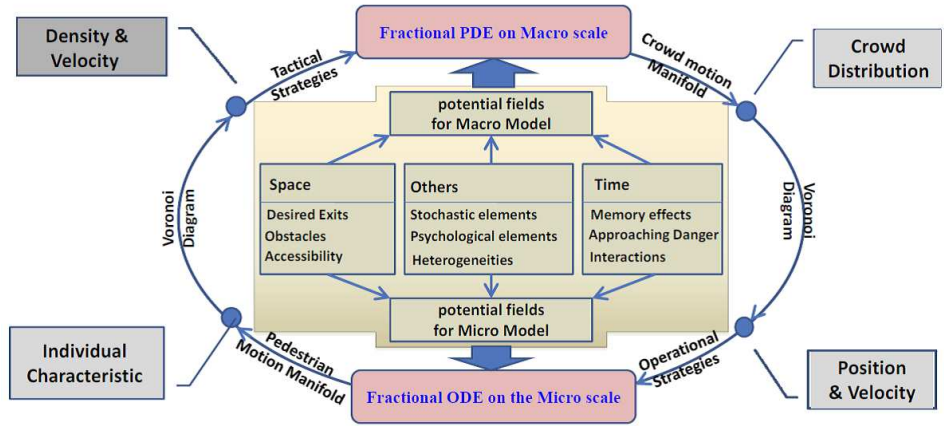}
\par\end{centering}

\centering{}\protect\protect\caption{\selectlanguage{english}%
\textit{\label{fig:Model-fractional} }\foreignlanguage{american}{\textit{Fractional
Model in Micro-scale and Macro-scale}}\selectlanguage{american}%
}
\end{figure*}

\par\end{center}

\end{itemize}

\subsection{Fractional Control of Crowd of Pedestrians}
\begin{itemize}
\item In microscopic level, the control of each pedestrian is focused on
analyzing the relationship between different interactions and collective
patterns where not only interactions based on topology but also interactions
based on short range and long range interactions will be considered.
Dynamics of crowd-pedestrian system under long range interactions
are closely related to fractional calculus and have an important effect
on generation of different collective patterns;

\begin{itemize}
\item This phenomenon observed from self-organized groups under different
interactions can be used to control the collective motion of crowds
where mobile Segways have been added into the group to tune the range
interactions among pedestrians to generate some desired collective
patterns;
\end{itemize}
\item In macroscopic level, fractional controllers based on fractional convection,
diffusion or both of them will be constructed for controlling of the
smoothing fluids (\ref{eq:my frac model}) for crowd of pedestrians
with fractional time/spatial orders;

\begin{itemize}
\item Due to the high densities in this scale, it is not easy to inject
control agents into the fluids. In our framework, Mobile Segways with
emergency personnels will be dispatched to control the inflow and
outflow of crowds from outside based on the theory of boundary control
to guarantee the smoothing evacuation of high-density crowds without
breakdowns;
\end{itemize}
\item In mescoscopic level, forward fractional convection-diffusion equations
and backward fractional Hamilton-Jacobi-Bellman (H-J-B) equations
are used for modeling the crowd pedestrian dynamics where the forward
part describes the evolution of crowds and the backward equation describes
the evolution of decision making process.

\begin{itemize}
\item Movement of next step will be prescribed based on the probability
distribution function as shown in backward part (\ref{eq:backward})
of the fractional HJB equations;
\item Fractional order controllers based on fractional diffusions or fractional
convections are adopted to realize the control or optimization of
the forward part (\ref{eq:forward}) of the fractional HJB equations;
\item Mean field games will play an important role in estimating neighbor\textquoteright s
influence and reducing burden of communication and computation;
\item Mobile Segways with global instructions can be used to guide or drive
the crowds through broadcasting or changing the structure of space
to control the velocity or flow of the crowds;
\end{itemize}
\end{itemize}
\textbf{Reasons of using the above control framework: }
\begin{itemize}
\item For crowds with low density, mobile Segways can be easily added into
the crowds and control of the crowd-pedestrian system can be realized
through interactions with just a few of them to realize control of
each pedestrian;
\item For crowds of high density, it is not wise to add mobile Segways into
the crowds again due to the high density. In this case, movements
of pedestrians are firmly constrained by the people around them. What
we try do is using more delicate controllers such as boundary control
to tune the inflow and outflow of people through Segways to realize
the smoothing fluids without breakdowns;
\item For crowds with medium density, there are some concerns that are not
easy to be solved in this scale. Besides heterogeneities and computation
burdens, the interactions between microscopic model and macroscopic
model are the main challenges. In other words, balancing between realization
of each pedestrian's target and desired evolution of the entire group
can be manipulated simultaneously in the framework of dynamic game
theory based on mean field.
\end{itemize}

\subsection{Platform for Verification }

Since real verifications of the theoretical results for evacuation
of crowd of pedestrians are not easy to conduct, a lot of simulation
results have been done in previous research. There are a lot of software
for the simulation research such as VISSIM, EXODUS, Simulex, PSCrowd,
PEDSIM, and VISWALK, et al.
\begin{itemize}
\item VISSIM is one powerful tool available for simulating multi-modal traffic
flows, including cars, buses, motorcycles, bicycles and pedestrians
and it is also a useful tool for the evaluation of various alternatives
based on transportation engineering. In VISSIM, movement of each pedestrian
is modeled by the social force model (\cite{HelbingDirk1995}) where
a total force resulting from the social, psychological, and physical
forces has been imposed. The forces that are influencing pedestrian's
motion are caused by his/her intentions to reach his destination as
well as by other pedestrians and obstacles. Thereby the other pedestrians
can have both attractive and a repulsive influences.
\item Due to the requirements of data processing and computations in modeling
and control of crowd of pedestrians, the simulation research is much
preferred to be realized using Matlab. The authors of this paper try
to study the modeling and control problem using the platform of DIFF-MAS
that has been developed in \cite{Liang2004} for simulating the measurements
and control of diffusion processes using Matlab script and Simulink.
Optimal placements of sensors and actuators, and structures for understanding
distributed networked actuation and sensing have been done on this
platform. Recently, this platform has been updated for measuring and
control of diffusive process of fractional order which is useful for
modeling and control crowds of pedestrians using calculus of fractional
order.\end{itemize}
\begin{rem}
\textcolor{black}{Considering unexpected or dangerous events in real-life
experiment, only initial theoretical results and some simple simulation
results have been done by the authors under the framework of fractional
order modeling and control of crowd of pedestrians. There are lots
of work left to show the effectiveness of this framework in future
especially from the aspect of control.}
\end{rem}

\section{Conclusion\label{sec:Conclusion}}

Recent advances in modeling and control of crowd of pedestrians have
been reviewed in the framework of Cyber-Physical Systems. In the first
part, not only modeling methods in micro-scale, macro-scale and mesco-scale
are reviewed but also possibilities of applying fractional calculus
are discussed; Due to challenges existing in the control problem,
only controllers in macro-scale and micro-scale have been surveyed
such as state feedback controllers, distributed controllers and robust
controllers. Initial considerations about control of crowd of pedestrians
are discussed in the framework of Cyber-Physical Systems. Long range
interactions are used to generate self-organized collective behavior
in the micro-scale; Fractional diffusion-convection controllers will
be constructed for crowd fluids with fractal time-spatial orders in
the macro-scale; Fractional Mean Field Games theory will be adopted
in modeling pedestrian's decision making process and control the fluids
of crowd with porosity in meso-scale. Multiple mobile Segways with
onboard emergency response personnels are employed to realize the
control of velocity and flux of crowd of pedestrians in different
scenarios.

A lot of problems are still open and needing cooperation of multi-disciplines
such as environmental design, engineering, and transportation. Potential
topics such as modeling the crowd of pedestrians with consideration
of \textbf{stochastic noise, psychological effects} and constructing
controllers under requirements of \textbf{scalability and robustness}
are both interesting topics and worthy of much efforts in future.
Another interesting and important topic is concerning the security
problem in the crowd of pedestrians where some recent results on security
of consensus problem such as \cite{He2013}, \cite{Liu2014} and \cite{Zhao2014}
are beneficial to solve of this problem.

\section*{Acknowledgment}

The authors gratefully acknowledge the Editors and the anonymous reviewers
for their valuable comments that helped to improve the quality of
this work.

\bibliographystyle{unsrt}
\bibliography{JAS,IEEEabrv,IEEEexample}

\end{document}